\documentstyle[12pt]{article}
\makeatletter
\if@twoside
\else \oddsidemargin 00pt \evensidemargin 00pt
\fi
\let\@eqnsel = \hfil

\expandafter\ifx\csname mathrm\endcsname\relax\def\mathrm#1{{\rm #1}}\fi
\@ifundefined{mathrm}{\def\mathrm#1{{\rm #1}}}{\relax}

\makeatother

\def\gtrsim{\ \rlap{\raise 2pt \hbox{$>$}}{\lower 2pt \hbox{$\sim$}}\ }
\def\lesssim{\ \rlap{\raise 2pt \hbox{$<$}}{\lower 2pt \hbox{$\sim$}}\ }

\def\ea{{et al.}}
\def\ib{{\it ibid.}}

\def\npb#1{Nucl. Phys. {\bf B #1}}
\def\plb#1{Phys. Lett. {\bf B #1}}
\def\prd#1{Phys. Rev. {\bf D #1}}
\def\prl#1{Phys. Rev. Lett. {\bf #1}}

\def\nature#1{Nature {\bf #1}}

\def\hb{\hfill\break}

\begin{document}
\thispagestyle{empty}
\null

\hfill INFN-FE 03-95

\hfill WIS-95/12/Mar-PH


\hfill March 1995

\vskip 2cm

\begin{center}
{\Large \bf      
Realistic SUSY Model with Four Fermion Families, \\
Natural R Parity and $\nu_\tau$ in the eV Range \\
\par} \vskip 2.em

{
{{\sc Zurab Berezhiani}$^{a,b}$ and {\sc Enrico Nardi}$^{c}$}
\\[1.5ex] 
{$^{a}$\it INFN, Sezione di Ferrara, 44100 Ferrara, Italy}\\
{$^{a}$\it Institute of Physics, Georgian Academy of Sciences, 380077
Tbilisi, Georgia}\\
{$^{c}$\it Department of Particle Physics, Weizmann Institute of
Science}\\
{\it P.O.B. 26, Rehovot, 76100 Israel}\\[2ex]
\par}
\end{center} \par
\vfil
\noindent
{\bf Abstract}
\par\noindent
We study an extension of the supersymmetric standard model with four
families and gauged horizontal symmetry $SU(4)_H$, in which R parity
automatically follows as a consequence of gauge invariance and of the
field content of the model. Beyond ensuring R parity conservation, the
horizontal symmetry allows for the dynamical generation of a hierarchical
pattern of fermion masses, without the need of any ad hoc choice of
small Yukawa couplings. The structure of the mass matrices implies that
the fourth family does not mix with the other three, and that the
$b',t',\tau',\nu'$ masses are all naturally in the 100 GeV range.
The model has other interesting phenomenological implications.
The scale of the horizontal symmetry breaking is constrained by
cosmological and astrophysical arguments to be $\sim 10^{11}$ GeV.
Then the late $b'$ decays could explain a magnitude of the isotropic
cosmic gamma flux. In addition, a lower bound of a few eV  is found
for the mass of the $\tau$-neutrino, which can thus provide a hot
component for the cosmological dark matter. Due to R parity conservation,
the lightest supersymmetric particle is stable and can provide the cold
component. The $\nu_e$ and $\nu_\mu$  masses can naturally be in the
correct ranges for a solution of the solar or atmospheric neutrino
problems via neutrino oscillations.
\par
\vskip 1cm
\noindent
\vfill
\noindent
--------------------------------------------\phantom{-} \hb
\noindent
Electronic mail:
berezhiani@fe.infn.it,
ftnardi@wizard.weizmann.ac.il
\null
\setcounter{page}{0}
\clearpage

{\large \bf 1. Introduction}
\vspace{4mm}

In the Standard Model (SM), Baryon (B) and Lepton (L) numbers are
conserved as a result of accidental global $U(1)_B$ and $U(1)_L$
symmetries that follow from the requirement of gauge invariance and
renormalizability.\footnote{These symmetries can be broken only by
non-renormalizable Planck scale operators like
$\frac{1}{M}ll\Phi\Phi$ \cite{non-ren-op}.
This term violates L and induces the neutrino Majorana masses of about
$10^{-5}$ eV, which could be relevant for the solar neutrino oscillations.
However, analogous terms violating baryon number are very
small to cause any observable effect. }
In the supersymmetric standard model (SSM) this
is not true anymore. In the superpotential, the renormalizable terms
\begin{equation}
l l e^c, \quad l q d^c, \quad  u^c d^c d^c.
\label{Rviol}
\end{equation}
($l$, $e^c$ and $q$, $u^c$, $d^c$ are
the lepton and quark superfields)  are allowed by the gauge symmetry,
which violate L and B.
In particular, the combination of first and third of these terms
 would lead to catastrophically fast proton decay
mediated by $d^c$-type squark exchange.
The relevant symmetry that ensures the absence of the terms (\ref{Rviol})
is called R parity, which is defined as $\rm R \equiv (-1)^{2J+3B+L}$,\
where J is the spin of the particle and  B(L) its baryon (lepton) number
\cite{R-parity}. R parity does not commute with supersymmetry (SUSY).
On the other hand, it is an automatic consequence of a $Z_2$ matter
parity under which the fermion superfields change the sign while the
`Higgs' ones $\Phi_{1,2}$ remain invariant.

An unsatisfactory feature of the SSM is that the $Z_2$ (or equivalently R)
parity conservation has to be imposed by hand.
The SUSY $SU(5)$ model does not differ much from the SSM,  since the
 gauge invariant term ${\bf 10} \, \bar{\bf 5}\,\bar{\bf 5}$,
involving the fermion superfields in the $\bf 10 +\bar {\bf 5}$
representation of $SU(5)$
leads again to the set of B and L violating couplings (\ref{Rviol}).
the  $SO(10)$ GUT offers an elegant solution to this problem:
the Higgs fields are generally assigned to vector
representations ({\bf 10}, {\bf 45}, {\bf 54}, {\bf 126} etc.)
while the fermion superfields are in the spinor representation {\bf
16}. While all the needed mass terms are allowed, the terms
in (\ref{Rviol}) would arise from ${\bf 16}^3$
which is forbidden by the gauge symmetry.
In other words, since  $SO(10)$ invariance allows for only pairs
of 16-plets, the theory has an automatic $Z_2$ matter parity
under which 16-plets change the sign whereas the superfields
in vector representations remain invariant.
However, this is not true anymore for the $SO(10)$
models in which symmetry breaking is triggered also
by the scalar components of superfields
belonging to the ${\bf 16}_{\Phi}+{\bf \overline{16}}_{\Phi}$
\cite{BabuBarr}.
In fact, after substituting the VEV $\langle {\bf 16}_{\Phi} \rangle$,
the couplings $\frac{1}{M}{\bf 16}^3{\bf 16}_{\Phi}$
lead again to the terms (\ref{Rviol}).
Since in these models the right handed neutrino masses are generated by
operators $\sim \frac{1}{M}{\bf 16}^2 {\bf 16}^2_{\Phi}$
of the same structure,
the ratio $\langle {\bf 16}_{\Phi} \rangle /M $ cannot be too
small, implying in turn that the magnitude of the resulting R-parity
violating terms  conflicts again  with the limits on the proton
lifetime. Thus R-parity conservation is not automatic  in
$SO(10)$ models with Higgses belonging to the ${\bf 16}_\Phi$,
and matter parity has to be imposed by hand in order to distinguish
the fermion 16-plets from the Higgs ones.

In this paper we wish to put forward the idea that R (or equivalently
$Z_2$) parity conservation can be naturally ensured in models based on
{\it gauged} non-Abelian horizontal symmetries.
Such models are particularly interesting since they can also explain
at least
qualitatively the observed pattern of fermion masses and mixing. Namely, the
structure of the fermion mass matrices can be related to the horizontal
symmetry breaking pattern, and the mass hierarchy between families
arises dynamically from certain hierarchies in this breaking
(see for example the models
\cite{horizontal,Khlopov} based on $SU(3)_H$ symmetry).
In this kind of models it is natural to assume that
the horizontal group $G_H$ acts only on the quark-lepton superfields,
while the Higgses $\Phi_{1,2}$ responsible for the electroweak
symmetry breaking are $G_H$-singlets.
Hence, independently of the choice of the vertical gauge group and/or
of the particular superfield assignments to its representations,
the Higgs and fermion superfields can be always distinguished
since they carry different horizontal quantum numbers.
This leads to the possibility of allowing the necessary mass
terms which are bilinear in the fermion superfields, while
forbidding the trilinear couplings (\ref{Rviol}).

Our task is now to find and classify the theories in which the
horizontal gauge group $G_H$  automatically forbids the unwanted
B and L violating terms in (\ref{Rviol}). Then
R parity (or equivalently $Z_2$ matter parity) will appear as an
automatic consequence of the horizontal gauge symmetry and of the
field content of the model. We demand that the models we are
interested in should satisfy the following list of
basic requirements:

$(i)$ In order to ensure a straightforward definition of the
horizontal gauge symmetry, all the fermion superfields
$q$, $l$, $ u^c$, $d^c$ and $e^c$ should be assigned to the same
representation of the non-Abelian horizontal group $G_H$. In other words,
we forbid $G_H$ singlet families.

$(ii)$ We require that the couplings in (\ref{Rviol}) are forbidden
by gauge invariance. Therefore $SO(3)_H$ and chiral $SU(3)_H$,
which are the only simple groups containing three dimensional
representations, are excluded since in both cases {\bf 3}$^3$
contains a gauge singlet. However we immediately notice that for
$SU(N)_H$ ($N>3$) the term $N^3$ does not contain gauge singlets.
Hence these groups represent a class of interesting candidates.

To have phenomenologically realistic theories, the
following additional constraints should be also imposed:

$(iii)$ In order to avoid the proliferation of Higgs doublets with
masses at the electroweak scale, the standard Higgs superfields
$\Phi_{1,2}$ must be $G_H$ singlets. The presence of several Higgs
doublets
would in fact spoil the natural suppression of Flavor Changing Neutral
Currents (FCNC). It would also destroy gauge coupling
unification, thus preventing any attempt to embed the model in some
vertical GUTs.

$(iv)$ The model should provide naturally
a realistic pattern of the fermion mass and mixing.
In this respect,  models based on  chiral $SU(3)_H$
\cite{horizontal,Khlopov} have proven to be quite successful in
relating the fermion mass hierarchy to a hierarchy in the
horizontal symmetry breaking VEVs, and in accounting for
the observed fermion mass and mixing pattern.
Even if chiral $SU(3)_H$ fails to satisfy condition $(ii)$,
this again suggests that chiral $SU(N)_H$ ($N > 3$) are promising
candidates.\footnote{ We note that vectorlike $SU(3)_H$ with $q,l$
transforming as {\bf 3} and $u^c,d^c,e^c$ as ${\bf \overline 3}$,
forbids the first two terms in (\ref{Rviol}),
which is enough to ensure proton stability. However it allows for
the $SU(3)_H$ invariant Yukawa terms $f f^c \Phi_{1,2}$. Then
the mass splitting between different families can be achieved
only at the price of {\it several} unnatural
fine tunings between horizontal non-invariant effective operators.
Thus this case can be hardly regarded as realistic.
(In addition  unification of the quarks and leptons within one
irreducible GUT multiplet would  clearly be  not  possible).
The same problem is encountered  for $SO(4)$ and $SO(5)$ with
4 dimensional representations. While the {\bf 4}$^3$ terms are forbidden
and condition $(ii)$ is fulfilled,
 the invariant Yukawa terms $f f^c \Phi_{1,2}$
are again allowed, and thus  condition $(iv)$ is not satisfied. }

$(v)$ A final strong condition is that R-parity breaking terms
should not appear even after $G_H$ breaking. In other words, no
effective couplings  which could generate the terms in
(\ref{Rviol}) after substituting the VEVs
$~\xi^k \rightarrow \langle\xi^k\rangle$
should be allowed by the $G_H$ symmetry.
In particular, this condition restrict the viable $SU(N)_H$
models to even $N$. Consider in fact $SU(N)_H$ with
the $f$ and $f^c$ fermion superfields assigned to the fundamental $N$
dimensional representation. The mass terms transform as $N\times N$ and
thus belong to two-index (symmetric and antisymmetric) representations.
In order to construct horizontal gauge invariant mass terms we can
take also the horizontal Higgses $\xi^k$  in two-index
representations. Then for $N$=4,6,\dots no invariants of the form
$N\times N \times N \times {\cal P}(\xi)$
(with ${\cal P}$ some polinomials in the horizontal fields
$\xi^k$) are allowed.
In contrast, for $N$ odd the totally antisymmetric
$\epsilon$ tensor allows to rewrite some combinations of Higgs fields
with an even number of free indices as tensors with an odd number of
free indices, suitable for generating effective
$N\times N\times N\times {\cal P}(\xi) $ gauge invariant terms.
After the breaking of the horizontal symmetry
($\xi^k\rightarrow\langle \xi^k \rangle$) these
effective operators
will then produce the R-violating terms in (\ref{Rviol}).

This brief analysis suggests that natural conservation of
R-parity could be achieved in models based on chiral horizontal
symmetries $SU(N)_H$ with $N$ even, under which the quark and lepton
superfields transform as  fundamental $N$-plets.
Hence the number of families must be extended to $N_f = 4,6,\dots $.
As is well known, the possibility of extra families with a light
neutrino is ruled out by the measured invisible decay width of
$Z$-boson,
however, sequential generations with heavy  neutrinos ($m > M_Z/2$)
are not excluded. On the other hand,
detailed studies \cite{fourth-fam} of
the effects of radiative corrections due to additional families show
that precise electroweak data are compatible with a fourth
family, while six families (which would be the next interesting case)
are ruled out. In addition, a dedicated analysis
showing the viability of SUSY models with four families
with respect to gauge coupling unification was presented in ref.
\cite{fourth-susy}. These results are relevant for our analysis,
since condition $(iii)$ ensures that the field content
in our $SU(4)_H$ model is the same than that of the four family SSM
of ref. \cite{fourth-susy}, up to some large energy scale where
the horizontal symmetry breaks down (see Sect. 2).
Hence, we conclude that if natural R parity conservation arises
due to some non-Abelian horizontal gauge symmetry, then
the theoretical and phenomenological constraints hint to models based
on the $SU(4)_H$ group,
on which we will concentrate in the rest of the paper.

\vspace{1cm}
{\large \bf 2. $SU(4)_H$ symmetry and its phenomenological implications}
\vspace{4mm}

Let us consider the standard SU(3)$\times$SU(2)$\times$U(1)
vertical gauge group, with local chiral $SU(4)_H$ horizontal symmetry
acting on four families of left chiral superfields
\begin{eqnarray}
f_\alpha: \quad & \quad
q_\alpha& = {u \choose  d}_\alpha\sim (3,2,1/6)_\alpha, \quad\quad\quad
l_\alpha = {\nu \choose  e}_\alpha \sim (1,2,-1/2)_\alpha \cr
f^c_\alpha: \quad &
u^c_\alpha& \sim (\bar{3},1,-2/3)_\alpha, ~~~
d^c_\alpha\sim (\bar{3},1,1/3)_\alpha,      \quad ~~
e^c_\alpha \sim (1,1,1)_\alpha
\label{sm-fields}
\end{eqnarray}
where each superfield is assigned to the fundamental {\bf 4}
representation ($\alpha=1,\dots 4$ is the $SU(4)_H$ index). With this
field content the horizontal $SU(4)_H$ is anomalous. In order to
cancel the horizontal anomaly we introduce the following superfields
which are vectorlike with respect to $SU(3)\times SU(2)\times U(1)$
and belong to the ${\bf \bar 4}$ of $SU(4)_H$:
\begin{eqnarray}
F^\alpha: ~~ & &
U^\alpha\! \sim\! (3,1,2/3)^\alpha,  ~~~
D^\alpha\! \sim\! (3,1,-1/3)^\alpha,  ~~~
E^\alpha\! \sim\! (1,1,-1)^\alpha      \cr
F_c^{\alpha}: ~~ & &
U_c^{\alpha}\! \sim\! (\bar{3},1,-2/3)^\alpha,  ~~
D_c^{\alpha}\! \sim\! (\bar{3},1,1/3)^\alpha,  ~~~
E_c^{\alpha}\! \sim\! (1,1,1)^\alpha,   ~~
N_c^\alpha\! \sim\! (1,1,0)^\alpha
\label{new-fields}
\end{eqnarray}
As we will see in short, these superfields turn out to be necessary
also for providing masses to the known fermions.

In the Higgs sector, we choose the standard Higgs doublet superfields
$\Phi_{1,2}(1,2,\mp1/2)$ to be singlets under $SU(4)_H$. In order to
break the horizontal symmetry and to generate the fermion masses, we
introduce also a set of `horizontal' Higgs superfields
$\xi^k_{\{\alpha\beta\}}$ ($k=1,2\dots$) transforming as the symmetric
{\bf 10}
representations of $SU(4)_H$ (clearly, they
must be $SU(3)\times SU(2)\times U(1)$ singlets).
Additional superfields $\bar{\xi}^k$
transforming as ${\bf \overline{10}}$ are also needed to render the
Higgsino sector free of chiral anomalies.
With this field content,
the most general Yukawa terms allowed by gauge invariance read as
\begin{equation}
W_F=g_f f_\alpha F_c^{\alpha} \, \Phi_{1(2)}
+ \sum_{k} h^k_F F^\alpha F_c^{\beta} \, \xi^k_{\alpha\beta}
+ \Lambda_f F^\alpha f^c_\alpha\,, ~~~~F=E,D,(U)
\label{W-fermions}
\end{equation}
\begin{equation}
W_N=g_\nu l_\alpha N_c^{\alpha} \, \Phi_{2}
+ \sum_{k} h^k_N N_c^\alpha N_c^{\beta} \, \xi^k_{\alpha\beta}
\label{W-neutrinos}
\end{equation}
\noindent
where the $g$'s and $h$'s are Yukawa couplings which we take to be
all $O(1)$. The last term in eq. (\ref{W-fermions}) is a gauge invariant
bilinear, and the $\Lambda_f$'s are some large mass parameters.
As already stated, no terms trilinear in the quark and lepton superfields
are allowed by the $SU(4)_H$ gauge symmetry, ensuring naturally the
absence of the B and L violating couplings (\ref{Rviol}).
We are facing here a situation analogous to the $SO(10)$ model,
since R-parity does not have to be imposed by hand,
but appears as an accidental symmetry
that follows from horizontal gauge invariance.
The terms (\ref{W-fermions}), (\ref{W-neutrinos}) are in fact invariant
with respect to the $Z_2$ parity under which the fermion
superfields $f,f^c,F,F_c$ change sign, while the Higgs superfields
$\Phi_{1,2}$ and $\xi$ stay invariant.
More in general, these have an automatic
global symmetry $U(1)_H$ under the following transformations:
\begin{equation}
f,f^c\to e^{i\omega}f,f^c,~~~
F,F^c\to e^{-i\omega}F,F^c,~~~
\xi^k \to e^{2i\omega}\xi^k,~~~
\Phi_{1,2}\to \Phi_{1,2}  ~~~
(\bar{\xi}^k \to e^{-2i\omega}\bar{\xi}^k)
\label{U1H}
\end{equation}
Its  $Z_2$ subgroup ($\omega\!=\!\pi$) remains unbroken even when
the scalars $\xi$ get non-zero VEVs. Clearly, this $Z_2$ matter
parity which ensures R parity conservation and hence proton stability
will be respected by all possible higher order Planck scale operators
as well.

The Yukawa terms (\ref{W-fermions}) lead to the so called
"universal seesaw" mechanism \cite{unisee,uniseesaw} for
generating the masses of the charged fermions,
 while the terms (\ref{W-neutrinos}) lead to
the ordinary seesaw mechanism for neutrino masses \cite{seesaw}.
After the horizontal scalars
$\xi^k$ develop non-zero VEVs, the extra fermions $F$ and $F_c$ of
eq. (\ref{new-fields}) acquire large masses through the second term
in eq. (\ref{W-fermions}). Then the first and third terms cause a
``seesaw'' mixing of the ordinary quarks and leptons $f,f^c$ with
the heavy ones.
As a result, in the base $(f,F)$ $(f^c,F_c)$, the $8\times8$ mass
matrix for the charged fermions $f=e,d,(u)$, $F=E,D,(U)$
reads
\begin{equation}
{\cal M}_f=
\pmatrix{0              & g_f v_{1(2)} \cr
         \Lambda_f   &  \hat{M}^F \cr} \,,  ~~~~~~~
\hat{M}^F_{\alpha\beta} =
\sum_{k} h^k_F \, \langle\xi^k_{\alpha\beta}\rangle
\label{charged-mass}
\end{equation}
where $v_{1,2}=\langle \tilde \Phi_{1,2}\rangle$
are the VEVs of the two electroweak Higgs doublets.
As for the neutral states, in the base $(\nu,N_c)$
the $8\times 8$ Majorana mass matrix has the form
\begin{equation}
{\cal M}_{\nu}=
\pmatrix{0              & g_\nu v_2 \cr
         g_\nu v_2  &  \hat{M}^N \cr} \,, ~~~~~~~
\hat{M}^N_{\alpha\beta} =
\sum_{k} h^k_N \, \langle\xi^k_{\alpha\beta}\rangle.
\label{neutral-mass}
\end{equation}
In contrast to the SM and most GUT models, in our picture the
fermion mass hierarchy is not generated by an {\em ad hoc} choice
of the Yukawa couplings. In fact
we can assume  all the Yukawa constants  to be
$O(1)$, close to the size of the gauge couplings.
As long as the off-diagonal blocks in eqs. (\ref{charged-mass}) and
(\ref{neutral-mass}) are flavour blind (unit) matrices, all the
informations on the fermion mass and mixing pattern is contained
in the heavy mass matrices $\hat{M}^F$.
The structure of the latter is determined by the different VEVs
$\langle \xi^k \rangle$ (modulo differences in the Yukawa constants
$h_F^k$), and the observed hierarchy of the light fermion masses is
ultimately determined by the hierarchy in the VEVs which break the
$SU(4)_H$ symmetry. In other words, the VEV pattern should provide
a step-by-step breaking of the chiral horizontal symmetry
\begin{equation}
SU(4)_H\times U(1)_H\stackrel{V_1}{\rightarrow}
SU(3)_H\times U(1)'_H\stackrel{V_2}{\rightarrow}
SU(2)_H\times U(1)''_H\stackrel{V_3}{\rightarrow} \tilde{U}(1).
\label{chain}
\end{equation}
The first breaking at the scale $V_1$ (given by the VEVs
$\langle\xi^k_{11}\rangle$)
provides the mass terms for the first heavy family $F_1$, the second
breaking (at $V_2\sim \langle\xi^k_{\alpha2}\rangle$, $\alpha\leq 2$)
for the second family $F_2$ and the third stage of breaking at
$V_3\sim \langle\xi^k_{\alpha3}\rangle$, $\alpha\leq 3$ generates the
mass of the $F_3$ state.
At this stage all the horizontal gauge bosons have acquired
large masses, the less massive  are the ones responsible for the
flavour-changing transitions between the fourth and the third families,
with masses $\sim V_3$.
The model also provides a natural possibility to obtain
together with three light families,
a heavy fourth, for which all the masses are
of the order of the electroweak scale.
In fact let us assume that all the VEVs of the type
$\langle \xi^k_{\alpha 4} \rangle $ are vanishing, so that a diagonal
global $\tilde{U}(1)$ subgroup of $SU(4)_H\times U(1)_H$, given
by the generator $\tilde{T}=\mbox{diag}(0,0,0,1)$, is left
unbroken.\footnote{A detailed analysis of the horizontal Higgs
potential will be given elsewhere.}
Then the $4\times 4$ mass matrices $\hat{M}^{F(N)}$ are
{\em rank}-3 matrices of the following form
\begin{equation}
\hat{M}^{F}=
\pmatrix{ M^{(3)}_F     &  0 \cr
               0  &  0    \cr}  ~, ~~~~~F=U,D,E,N
\label{degenerated}
\end{equation}
where the $3\times 3$ blocks $M^{(3)}_F$ contain non-zero entries,
given by the VEVs of the horizontal scalars.
Clearly, the residual global symmetry $\tilde{U}(1)$ ensures
naturally the heaviness of the fourth family, since
there is no seesaw mechanism for the corresponding fermions.
The right-handed components of the fields
$f_4=b',t',\tau',\nu'$ are actually the $F_c^4$ states, whereas the
$f^c_4$ form with the $F_4$ superheavy particles of mass $\Lambda_f$.
{}From eqs. (\ref{charged-mass}) and (\ref{neutral-mass}) we obtain
\begin{equation}
m_{b',\tau'}=g_{d,e} v_1,~~~~~m_{t',\nu'}=g_{u,\nu} v_2.
\label{fourth-mass}
\end{equation}
Since the Yukawa couplings are all $O(1)$, for moderate values
of $\tan\beta=v_2/v_1$ all the masses in (\ref{fourth-mass}) are
of the order of electroweak scale $\sim$ 100 GeV. On the other hand,
it is apparent from (\ref{charged-mass}) that the fermions of the first
three families will acquire their masses through a seesaw mixing with the
superheavy $F$ fermions. Their masses will then be  suppressed
from the electroweak scale down to the observed values, provided that
$M_F^{(3)}> \Lambda_f$.
Namely, after decoupling the heavy states\footnote{ Below the scale
$V_H=V_3$ our theory simply reduces to the SSM with four families.
Therefore, for $V_H$ large enough
all FCNC phenomena related to the horizontal gauge or Higgs bosons and
to the mixing with weak isosinglet heavy fermions are strongly suppressed. }
the $3\times 3$
mass matrices of the light charged fermions $f=e,d,(u)$ are
$m_f^{(3)} = g_f \mu_f (M_F^{(3)})^{-1} v_{1(2)} $
while for the $3\times 3$ Majorana mass matrix for the light neutrinos
we obtain
$m_\nu^{(3)} = (M_N^{(3)})^{-1} (g_\nu v_2)^2.$
Hence through the seesaw mechanism the horizontal VEV pattern (\ref{chain})
is reflected in the observed pattern of fermion masses in an inverted
way \cite{unisee} (see also \cite{Rattazzi}).
 Modulo the different $O(1)$ Yukawa constants $h^k_F$
the heavy mass eigenstates of $M^{(3)}_F$ reflect the hierarchy
$V_1\gg V_2\gg V_3$ of the $SU(4)_H$ symmetry breaking.
On the other hand, the ordinary quark and lepton masses are inversely
proportional to the masses of their heavy partners. For example,
for the charged lepton and the neutrino masses
we have\footnote{Since in the following we
deal with order of magnitude estimates, we neglect
the different $O(1)$ factors accounting for the renormalization group
running of masses from the horizontal scale to lower energies. }
\begin{equation}
m_{e,\mu,\tau}\simeq \frac{\Lambda_e m_{\tau^\prime}}{h^{1,2,3}_E V_{1,2,3}}
< m_{\tau^\prime}\,,~~~~~~
m_{\nu_e,\nu_\mu,\nu_\tau} \simeq \frac{m_{\nu'}^2}
{ h^{1,2,3}_N V_{1,2,3} } \ll m_{\nu'}
\label{seesaw}
\end{equation}
The quark masses are given by expressions analogous to that of
charged leptons. The CKM matrix of mixing between the first three
families is also determined by the structure of the matrices $M^{(3)}_F$,
and arises from the presence of  non-diagonal VEVs
$\langle \xi^k_{\alpha\beta} \rangle$ ($\alpha,\beta =1,2,3$).
The fact that the $b$ and $\tau$ masses are of order
a few GeV, implies that
the masses of the corresponding heavy states $h^3_DV_3$ and $h^3_EV_3$
are not much larger (say, within one or two orders of magnitude)
than the mass scale $\Lambda_{d,e}$. As for the top quark, the value of
its mass $m_t\gtrsim 150\,$GeV requires $h^3_U V_3$
$\sim \Lambda_u$. In this case corrections to the seesaw formula
should be taken into account in relating $m_t$ to the heavy scales
(see e.g. \cite{Rattazzi}).

As we stated above, the residual $\tilde{U}(1)$ symmetry implies
that fourth neutrino $\nu'$ is a Dirac particle with mass
$\sim 100\,$GeV, while three neutrinos get small Majorana masses
given by eq. (\ref{seesaw}).
Therefore, the neutrino mass hierarchy is expected to be approximately
the same as the hierarchy between the quarks or the charged leptons:
\begin{equation}
m_{\nu_e}: m_{\nu_\mu}:m_{\nu_\tau} \sim m_u : m_c :m_t ~~~~~
{\rm or} ~~~~~\sim ~m_e:m_\mu : m_\tau.
\label{neutrino-hierarchy}
\end{equation}

On the experimental side,
the firmest constraints on the masses of any new sequential fermion,
quark or lepton, have been set at LEP: $ m_f \gtrsim M_Z/2$.
This indeed represents the best constraint on
$m_{\tau^\prime}$ and $m_{\nu^\prime}$.
Searches for new quarks at the TEVATRON collider could in principle
give much better bounds for
$m_{t^\prime}$ and $m_{b^\prime}$\cite{bprime-limits}.
However, let us note that the structure of the heavy mass matrix
(\ref{degenerated}) implies that the fourth family is unmixed with the
three lighter ones. Hence the usual signatures, as for example
$b^\prime \to c,u$, that have been used to set the limits on new
sequential quarks \cite{bprime-limits} do not occur in our case.
In the absence of a detailed experimental analysis of the unmixed case,
the only reliable limit is again the LEP one also for the new quarks.
Hence we can safely conclude that the predictions in (\ref{fourth-mass})
are by no means in conflict with the existing experimental limits.
However, it is clear that for the masses of the fourth family fermions
not much room is left. The allowed parameter space is strongly
constrained by the CDF measurement of the top mass,
$m_t=174\pm 10\pm 13\,$GeV \cite{CDF},
by the precision tests of the SM which do not
leave much space for additional sizeable radiative
corrections as would be induced by a too large
$m_{t^\prime}$--$m_{b^\prime}$ splitting, and by renormalization group
analysis of the Yukawa couplings, much in the spirit
of ref. \cite{fourth-susy}.
In particular,
the universal seesaw mechanism
 implies $m_{t^\prime} \geq  m_t$, and most likely
$m_{t^\prime} \geq m_{b^\prime}$.
Then, according to \cite{fourth-susy}, for
$m_{t^\prime} \geq  m_t > 150\,$GeV the consistency of
the model implies not too large masses for the
fourth family fermions.
For the low values of $\tan\beta$ we are interested in
(e.g. $\tan\beta \sim 2$), the maximal values allowed are about
$m_{b^\prime}\sim 100\,$GeV and
$m_{\tau^\prime,\nu^\prime} \sim 50\,$GeV,
that is within the reach of LEP II.

\vspace{1cm}
{\large \bf 3. Astrophysical consequences of the model }
\vspace{4mm}

Indeed the presence of a fourth family of fermions
constitutes the most compelling prediction of our model,
so we will now address some issues regarding these new states.
The unbroken diagonal $\tilde{U}(1)$ subgroup of $SU(4)_H\times U(1)_H$
implies that the fourth family is unmixed with the three lighter ones.
We assume that the lightest member of the fourth generation is
the neutral one $\nu^\prime$, as is also suggested by the analysis
of ref. \cite{fourth-susy}. For simplicity we also assume that
$m_{t^\prime} > m_{b^\prime}$.  Then  $b^\prime$ and $\nu^\prime$
are stable with respect to electroweak interactions.
A stable $\nu^\prime$ with mass in the 100 GeV
range is cosmologically safe,
the contribution $\rho_{\nu'}$ to the present cosmological
density would not exceed the critical density $\rho_c$
even in the presence of a
$\nu^\prime$-$\overline{\nu^\prime}$ primordial asymmetry
of the order of the baryon asymmetry.
However, as we will argue in the following,
in our model no sizeable asymmetry has to be expected
for the fourth family fermions.

In contrast, the existence of stable heavy quarks carrying colour
and electric charge would constitute serious problem,
since it will conflict with the constraints arising from superheavy
element searches \cite{heavy}. Indeed, the stable $b^\prime$ would
behave essentially as $d$ quarks, hadronising
into heavy stable
mesons $\bar b^\prime u$ and `nucleons' $b^\prime u d $
\cite{heavy}
giving rise to heavy hydrogen-like `isotopes' with masses
$\sim 100$ GeV. The existing experimental limits on this kind of
isotopes are extremely tight. For $m_{b^\prime}< 1$ TeV
the limit on the $b^\prime$ abundance relative to normal hydrogen
is $n_{b^\prime}/n_B < 10^{-28}$  \cite{heavy-H}.

However, the exchange of the $SU(4)_H$ gauge bosons $Z_H$ would
allow the heavy quark to decay, dominantly  through the channel
$b^\prime \to b \bar{\nu}_{\tau} \nu^\prime $, with a lifetime
\begin{equation}
\tau_{b^\prime}\simeq \left(\frac{V_H}{v}\right)^4
\left(\frac{m_\mu}{m_{b^\prime}}\right)^5
\tau_\mu =
\left(\frac{V_H}{10^{12}~ \mbox{GeV}}\right)^4
\left(\frac{150~\mbox{GeV}}{m_{b^\prime}}\right)^5
\cdot 4\cdot 10^{17}~\mbox{s}
\label{lifetime}
\end{equation}
where $V_H=V_3$ is the lowest scale in the horizontal symmetry
breaking (see eq. (\ref{chain})), $v=174\,$GeV is the electroweak scale
and $\tau_\mu = \tau(\mu\rightarrow e\bar{\nu}_e \nu_\mu)$
is the muon lifetime.
We can use cosmological arguments, together with the
experimental limits on searches for heavy isotopes, to put an upper
bound on $\tau_{b^\prime}$, which in turn will translate in an upper
limit on $V_H$.
Taking into account the finite lifetime of the heavy quarks,
their present number abundance relative to baryons is
$\,n_{b^\prime}/n_B = r_0 \exp({-t_0/\tau_{b^\prime}}) \leq 10^{-28}$,
where $r_0=(n_{b^\prime}/n_B)_0$ is the relic abundance for
stable $b^\prime$,  $h=0.5-1$ is the Hubble parameter in units of
$100\,$Km s$^{-1}\,$Mpc$^{-1}$, and $t_0\simeq 2\cdot 10^{17}h^{-1}\,$s
is the present age of the Universe (we assume matter dominated
expansion of the
Universe, and $\Omega=1$ as it is motivated by inflation).
{}From this equation we get an upper limit on the $b^\prime$ lifetime
\begin{equation}
\tau_{b^\prime}\leq 3.1 \cdot 10^{15} h^{-1}
(1+ 0.036\,\lg r_0)^{-1}\, \mbox{s}.
\label{tau}
\end{equation}
Due
to many theoretical uncertainties related to the actual annihilation
cross section for the $b^\prime$,
it is not possible to compute precisely the value of $r_0$,.
However, an estimate of the relic abundance
of heavy stable $d$-type quarks has been given in \cite{heavy}.
Under the assumption that there is no cosmological
 $b^\prime$ and  $\bar{b}^\prime$  asymmetry,  it was found that for
$m_{b^\prime} \sim 150\,$GeV
the energy density of these relics, relative to critical density
(namely $\Omega_{b^\prime} h^2$ )
could range from $10^{-9}$  to
$10^{-4}$ (smaller values are obtained for lighter $b^\prime$
masses). The lower limit corresponds to the case when the relic
density is determined by  the annihilation after the QCD phase
transition, and it was obtained by taking as an upper bound
on the annihilation cross section the geometrical cross section
($\sigma_0\sim 100\,$mb ).
The upper limit was obtained under the opposite assumption, namely that
annihilation after confinement is negligible, and that
the relic density is essentially determined by the QCD annihilation
cross section for free quarks.
Then the ratio of the $b^\prime$ to baryon number densities
$r_0 = (n_{b^\prime}/n_B)_0 =
\Omega_{b^\prime} /\Omega_B \cdot m_B/m_{b^\prime}$
lies in the range $3\cdot 10^{-10} < r_0 < 3\cdot 10^{-5}$
where we have taken $\Omega_B \sim 0.02$ as suggested
by nucleosynthesis estimates.
As is discussed in \cite{heavy}, the most
reasonable assumption is that the relevant annihilation process happens
after confinement, however with a cross section much smaller than the
geometrical one, giving $r_0 \sim 10^{-7}-10^{-8}$.
Clearly, in the presence of a sizeable baryon asymmetry in the fourth
family sector, the relic abundance of the heavy $b^\prime$ quarks
would be some orders of magnitude larger than the quoted estimates,
up to $r_0\sim 1$.

As we see, the bound (\ref{tau}) very weakly depends on the initial
$b^\prime$ abundance. Even if we
let $r_0$ range between $1-10^{-10}$, by taking
$h=0.5$ we obtain $\tau_{b^\prime}\leq 6\cdot 10^{15}-10^{16}\,$s.
On the other hand, according to eq. (\ref{lifetime}),
this bound translates into an upper limit
\begin{equation}
V_H \leq\, \frac{ (m_{b^\prime}/150~\mbox{GeV})^{5/4} }{ h^{1/4}
(1+ 0.036\,\lg r_0)^{1/4} }\cdot 3\cdot 10^{11} ~\mbox{GeV}
\leq  4 \cdot 10^{11} ~\mbox{GeV}
\label{V_H}
\end{equation}
where we have taken the conservative upper bound
$m_{b^\prime}<150\,$GeV  from the analysis \cite{fourth-susy}.
For more realistic values $m_{b^\prime}\simeq 100\,$GeV
\cite{fourth-susy} we get $V_H\leq 2.4\cdot 10^{11}\,$GeV.
According to eq. (\ref{seesaw}), this upper limit on the scale $V_H=V_3$
together with the experimental limit $m_{\nu^\prime} \geq M_Z/2$
translates into a lower bound on the $\tau$-neutrino mass:
\begin{equation}
m_{\nu_\tau}\simeq h_N^{-1}\,\frac{m_{\nu'}^2}{V_H}\gtrsim (1-10)~\mbox{eV}
\label{nutau}
\end{equation}
where in the numerical estimate we have taken into account the $O(1)$
uncertainties in the Yukawa coupling $h_N$ (for perturbativity we
have to assume $h_N \leq 3$).
On the other hand, the most conservative cosmological upper bound on
the light stable
neutrino masses $m_{\nu_\tau} \leq 92\,\Omega\, h^2$ eV \cite{GZ}
provides a lower bound on $V_H$:
\begin{equation}
V_H \geq \frac{h_N^{-1} }{4\Omega h^2} \left(\frac{m_Z^2}{92\,\mbox{eV}}
\right)\gtrsim (0.3-1)\cdot 10^{11}~ \mbox{GeV}
\label{univ_age}
\end{equation}

More stringent limits on $r_0$ and $\tau_{b^\prime}$ can be derived
by considering that the late decay of the $b^\prime$ can cause a
significant contribution to observed cosmic ray fluxes, in particular
to the isotropic diffuse gamma-ray background \cite{Berez}.
Indeed, at the moment of decay,
the $b^\prime$ quarks are bounded within colorless hadrons like
$b^\prime u d$ or $\overline{b^\prime} u$ \cite{heavy}.
Then the decay $b^\prime \to b \overline{\nu}_{\tau} \nu^\prime$ will
produce an hadronic jet with the $b$ quark being the leading
particle and an excitation energy $E_0\simeq \frac{1}{3}m_{b^\prime}$.
The fragmentation of this jet produces $\pi^0$, $\eta$ etc., with
the subsequent radiative decay resulting in a specific photon spectrum, and
the number of photons produced is directly proportional
to $r_0$. In order to estimate the flux at the present era,
the redshift in the photon energies has to be taken into account as well.
As long as the decay happens at the matter dominated epoch, and the small
amount of relativistic decay products does not affect sensibly the
Universe expansion rate, we have
$1+z=(t_0/\tau_{b^\prime})^{2/3}\sim 10-20$
for the values of $\tau_{b^\prime}$ estimated above.
To compute the value of the isotropic cosmological gamma-flux
we need to know what fraction of the
jet energy $E_0$ is taken by the photons and what is the energy
spectrum.
The photon spectra produced in jet hadronization
for different leading particles were computed,
using a Monte Carlo simulation program,
in ref. \cite{Salati}.
These spectra exhibit a remarkable scaling property
in terms of the variable $x=E_\gamma/E_0$,
and in the case the leading particle is a $b$ quark,
the photons carry away about 25 percent of the initial jet energy.
Using the results of ref. \cite{Salati}
we have estimated the present value of the gamma-flux
$d\Phi_\gamma/dE_\gamma$ as a function of $r_0$,
and we have compared it with the existing
observational limits (see \cite{Berezinsky} and references therein).
For example, for $E_\gamma=100$ MeV
the experimental bound is $d\Phi_\gamma/dE_\gamma \lesssim
10^{-7}\,$cm$^{-2}\,$s$^{-1}\,$sr$^{-1}\,$MeV$^{-1}$ and
the cosmic gamma-flux produced from $b^\prime$ decays
at $z=10-20$ saturates this bound for $r_0\sim 10^{-7}$.\footnote{
Substantially larger $r_0$ would require much larger redshift, and
hence much smaller $\tau_{b^\prime}$. However, the lower bound
(\ref{univ_age}) on the breaking scale $V_H$ excludes much
smaller lifetimes. }
Thus, for the preferred values $r_0\sim 10^{-7}-10^{-8}$ \cite{heavy}
our model is consistent with  with the observational data on the
isotropic gamma-background, and can even provide an
appealing explanation of its origin.

The previous analysis implies that $r_0$ should be rather small.
In particular this excludes any
sizeable baryon asymmetry between $b^\prime$ and $\bar{b}^\prime$.
Then the possible baryogenesis
mechanisms applicable to our model are severely constrained.
Even if a baryon asymmetry in the fourth family
is hardly expected, since there is no mixing with the other
three families, and hence no source of CP violation,
the high rate of sphaleron processes before the electroweak
phase transition \cite{KRS} would immediately
redistribute any baryon asymmetry present in the first three
families to the fourth one. Therefore, no mechanism
is acceptable which generates the baryon asymmetry before the
sphaleron effects are switched off.\footnote{In principle,
in our model the baryogenesis with non-zero B$-$L could occur
due to CP violation effects in out-of-equilibrium decays
$N^c \to l+ \Phi$ of the heavy right-handed neutrino \cite{FY} (for
the viability of this mechanism in the SUSY case see
ref. \cite{Olive}), or in the decays of $SU(4)_H$ gauge or scalar
bosons. Then sphaleron effects would immediately transfer the
produced net lepton number into a baryon asymmetry also in the
fourth family sector. However, our model naturally avoids the
possibility of such a lepto-baryogenesis.
As it was shown in ref. \cite{Olive}, the large scale density
fluctuations hinted by the COBE measurements require rather
low inflationary reheat temperature ($T_R\sim 10^8\,$GeV) and
correspondingly low inflaton mass ($m_\eta\sim 10^{11}\,$GeV).
Then for $V_H\sim 10^{11}\,$GeV
the masses of {\em all} right-handed neutrinos and horizontal bosons
are $\gtrsim m_\eta$,
and therefore they cannot be produced after inflation. }
In the context of our model, the most appealing scenario
is to assume that baryon asymmetry is generated
at the electroweak (first order) phase transition, as a result
of CP violation in the dynamics of quarks or leptons interacting
with the walls of the
expanding bubbles of the broken phase \cite{Misha}. Outside of
the bubbles the electroweak symmetry is unbroken, quarks are massless
and the rate of the fermion number violation due to
sphaleron transitions greatly exceeds the Universe expansion rate.
Inside the bubbles the quarks are massive due to non-zero VEVs of the
Higgs fields, the sphaleron processes are strongly suppressed
and fermion number is effectively conserved. Baryon asymmetry
inside the bubbles could be produced (and maintained) due
to CP violating effects, as a difference between the quark and
anti-quark fluxes penetrating the walls from the unbroken phase to the
broken one. This will affect only the first three family
fermions. Since the fourth family is unmixed and has no CP violation,
no baryon excess is expected in this sector.
Although the viability of such a baryogenesis
in the SM is still disputed in the literature, in the context of SSM it
could be more effective and sufficient for providing the observed baryon
asymmetry. Clearly this topic deserves additional special considerations.


Few remarks about the neutrino masses are now in order.
For the $\tau$-neutrino mass our model predicts the  lower bound
eq. (\ref{nutau}) of a few eVs ,
and according to (\ref{neutrino-hierarchy})
$\nu_\mu$ and $\nu_e$ are expected to have much smaller masses.
A $\nu_\tau$ with mass in the range $1-10$ eV will give a sizeable
contribution to the cosmological energy density as a hot dark matter
(HDM) component.
We remind here that the COBE measurements of the
cosmic microwave background anisotropy, together with other data on
the density distribution of the Universe at all distance scales
(galaxy-galaxy angular correlations, correlations of galactic
clusters, etc.), can all be fit by assuming some HDM admixture
to the dominant CDM component \cite{Shafi}. The best fits hint to a
neutrino mass $m_{\nu_\tau} \sim 5-7$ eV \cite{Primack}
which does appear naturally in our model.
As for the CDM itself, in our R parity conserving SUSY model it can
be naturally provided by the lightest supersymmetric particle (LSP),
presumably a neutralino.

As we commented earlier, the neutrino mass hierarchy should be
qualitatively the same as that for the charged quarks and leptons.
However, the spread in the Yukawa coupling constants $h_F$
does not allow to put severe limits on the other neutrino masses.
For example, by taking $m_{\nu_\mu}/m_{\nu_\tau} \sim m_c/m_t$, as
is suggested by the first estimate in eq. (\ref{neutrino-hierarchy}),
one obtains $m_{\nu_\mu}\sim (2-5)\cdot 10^{-3}\,$eV. This range
corresponds to the
MSW solution of the solar neutrino problem
\cite{MSW} via $\nu_e\to \nu_\mu$ oscillations.
Alternatively, if we had to attempt an explanation
of the deficit of the atmospheric $\nu_\mu$ via $\nu_\mu\to \nu_e$
oscillations (for a recent analysis, see \cite{atm-new}),
we would need $m_{\nu_\mu}\sim 0.1$ eV which is compatible with the
second estimate in eq. (\ref{neutrino-hierarchy}).
Obviously the MSW explanation to the solar
neutrino deficit would not be viable in this latter case.

\vspace{1cm}
{\large \bf 4. Conclusions }
\vspace{4mm}

In this paper we have put forward the idea that natural
conservation of R parity in SUSY models can be guaranteed
in the presence of a suitable horizontal gauge symmetry,
so that an accidental $Z_2$ matter parity (equivalent to R parity)
automatically follows from gauge invariance and the field
content of the model.
On theoretical and phenomenological grounds, we have
identified $SU(4)_H$ as the only viable horizontal gauge group for
implementing this idea. It implies existence of four fermion families.
We have suggested a realistic SUSY model based on the SM vertical gauge
group $SU(3)\times SU(2)\times U(1)$ with an $SU(4)_H$
anomaly free horizontal gauge symmetry. It
leads to a particular form of the fermion mass matrices rendering
all the fourth family fermions
naturally in the 100 GeV range and unmixed with the first three
families. This is due to the global $\tilde{U}(1)$ symmetry
which remains unbroken and in fact represents the conserved fermion
number of the fourth family.\footnote{
Let us remark that actually this is a rather general statement
which applies not only to our particular SUSY model designed for
automatic R parity: if the fourth family fermions
will be indeed discovered in future colliders,
it is not difficult to convince oneself
that the only natural way of their accomodation is to
to prescribe them the global conserved "fourth flavour" number
which renders the fourth neutrino to be a Dirac particle with
$O(M_W)$ mass.  }
As for the masses of the light three families,
our model leads to "seesaw" suppression of their magnitude
from the electroweak scale down to the observed values.
This suppression is achieved dynamically,
without the need of any tuning
of the Yukawa couplings which have been assumed to be all $O(1)$.
By means of cosmological and astrophysical arguments,
we have constrained rather precisely the scale $V_H$
at which the horizontal gauge symmetry is completely broken,
obtaining a narrow window around $10^{11}\,$GeV.
The upper bound on the scale $V_H$ sets a lower limit
on the $\tau$ neutrino mass of about few eV.
Hence our model naturally provides cosmological HDM
in the form of $\nu_\tau$ and, due to R parity conservation,
also CDM in the form of stable LSP.
Since in our scheme conservation of R-parity is
ensured by the horizontal gauge symmetry independently
of the particular choice for the vertical gauge group,
it would be interesting to extend  the present analysis
to phenomenologically appealing GUT models,
such as $SU(5)$ or $E_6$, for which R-parity
conservation is not automatic.

\vspace{1cm}
{\large \bf Acknowledgements}
\vspace{0.4cm}

\noindent
It is a pleasure to thank Venya Berezinsky, Sasha Dolgov
and Misha Shaposhnikov for illuminating
discussions and useful comments.

\vspace{1cm}


\end{document}